# Oxygen out-diffusion in REBCO coated conductor due to heating

Jun Lu, Yan Xin, Brent Jarvis, and Hongyu Bai

Magnet science and technology, National high magnetic field laboratory, Tallahassee, FL32310

**Abstract**

Rare earth barium copper oxide (REBCO) coated conductor has emerged as one of the high $Tc$ superconductors suitable for future ultrahigh field superconducting magnet applications. In the design and fabrication of such ultrahigh field REBCO magnets, it is essential to understand the behavior of REBCO coated conductor. The effect of heating on the properties of commercial REBCO coated conductors is very important for many practical reasons. Nevertheless, a comprehensive study on this effect have not yet been presented in the published literature. This work studies a commercial REBCO coated conductor heat-treated at temperatures between 175 °C and 300 °C for various durations. Critical current and lap joint resistivity were measured at 77 K and 4.2 K for the heat-treated samples. We found that critical current degrades with heat treatment time and temperature. This degradation can be described by a one-dimensional oxygen out-diffusion model with a diffusion coefficient of $D$ = 2.5 x $10^{-6}$ exp (-1.17 $eV/kT$) m$^2$/s. The heat treatment also causes appreciable increase in joint resistivity. Comprehensive structural and chemical analyses were performed on Cu/Ag/RECBO interfaces by transmission electron microscopy (TEM). Our electron energy loss spectroscopy (EELS) study provided direct evidence of oxygen deficiency in the heat treated REBCO samples. In addition, it is found that the oxygen diffused out of the REBCO layer forms mostly $Cu_2O$ at both Ag/REBCO and Cu/Ag interfaces. $Cu_2O$ is also observed at grain boundaries of the Ag layer. The oxygen out-diffusion model proposed in this work is used to predict REBCO thermal degradation in several engineering scenarios.

**Keywords**: REBCO, oxygen diffusion, thermal degradation, critical current, joint resistivity,

I.      Introduction

Rare earth barium copper oxide (REBCO) coated conductor has been developing rapidly since the last decade. It has achieved very high critical current density at liquid helium temperatures in very high magnetic fields. This allows it to be used as the preferred conductor for ultrahigh field superconducting magnets development [1]-[3]. REBCO coated conductor, however, is a relatively new technology, many aspects of its properties are still being explored by researchers in the superconducting magnet community. For ultrahigh field magnet applications in particular, the properties of the REBCO conductors have to be fully explored. The fundamental physics and materials science of these properties have to be well understood as well. One of these important properties is the effect of heating on REBCO's critical current ($I_c$) and lap joint resistivity.

Heating is inevitable during REBCO coil fabrication and operation. For example, heating is required in the soldering processes, such as lamination of copper stabilizer or fabrication of solder joints and terminals. While during magnet quenches the hot spots in REBCO can reach considerably high temperatures although only for a short period of time. Similarly, in power applications such as fault current limiters, fault current much higher than critical current may occur for a short period of time heating REBCO conductor considerably [4].

The experimental results in the literature indicated that $I_c$ degrades with increasing heating temperature and time [5]-[7]. This suggests that heating of REBCO should be minimized whenever possible. In some cases, however, controlled $I_c$ degradation could be beneficial. This is when $I_c$ needs to be slightly reduced from its nominal production value in order to minimize the electromagnetic stress due to screening current in various magnet coils. Some research and development have already pointed to this direction [8],[9].

Meanwhile the practical piece-length of commercial REBCO tapes is typically limited to about one hundred meters at present, large number of solder joints are needed in a practical ultrahigh field REBCO magnet. Therefore, having low joint resistivity is essential to the energy efficiency. Since the joint resistivity is dominated by the interfacial resistivity between the REBCO and the silver layer [11], it is conceivable that heating changes the chemistry of the REBCO layer especially near the REBCO/Ag interface which in turn changes the joint resistivity. Understanding the heating effect on joint resistivity could help us to determine the maximum allowable joint soldering temperature and time. The study in this area, however, has not been reported.

In this work, we systematically study the effect of heating on both $I_c$ and joint resistivity of REBCO coated conductor made by SuperPower Inc. The data are analyzed by using a simple oxygen diffusion model, which is then used to predict the $I_c$ and joint resistivity of REBCO as functions of heat treatment time and temperature. In addition, a comprehensive transmission electron microscopy study provided direct evidence of oxygen out-diffusion in heat-treated samples.

## II. Experiment

Two types of REBCO tapes are used in this work. Both are 4 mm wide made by SuperPower. One is SCS4050-AP with 20 μm thick copper stabilizer on each side. The other is SF4050-AP without Cu stabilizer (silver finish only). Fig. 1(a) is an optical micrograph of a cross-section of SCS4050-AP tape. From the 50 μm Hastelloy substrate up, there is a thin metal oxide buffer layer, a REBCO layer of about 1 μm (dark contrast), a silver layer of less than 1 μm (light contrast), and a 20 μm Cu stabilizer layer. Scanning transmission electron microscopy (STEM) images of more details of two locations in Fig. 1(a) is shown in Fig. 1(b) and 1(c) respectively.

The REBCO tapes were cut to 80 mm long samples for heat treatments. Heat treatment were performed in either air, pure argon or vacuum at temperatures between 175 °C and 300 °C in a quartz tube furnace. Air and pure argon were at ambient pressure. The vacuum was better than 1 x $10^{-4}$ millibar.

The transport critical current of heat-treated samples was measured in liquid nitrogen and self-field by using the four-probe method. The distance between voltage taps was 40 mm. The criterion for determining critical current is 1 µV/cm. The magnetization critical current was measured at 4.2 K in 8.5 T magnetic field by a vibrating sample magnetometer in a physical property measurement system made by Quantum Design. In this measurement, the applied magnetic field was perpendicular to REBCO's *ab* plane; and the typical sample size was 4 x 7 $mm^2$. A formula $M = (1/2)J_cw(1-w/3l)$ [10] is used to convert magnetization $M$ to critical current density $J_c$, where $w$ and $l$ are sample width and length, respectively.

For joint resistivity studies, lap joints were made from samples heat treated in argon. This is to minimize the effect of surface oxidation on joint resistivity. In order to further minimize the surface effect, the sample surface was lightly polished by Scotch-brite abrasive pads before the lap joint was soldered. The lap joints are 25 mm long, soldered by Pb37Sn63 solder with the help of a dedicated soldering fixture [11] at about 210 °C for about 1 minute. The joint resistivity was measured using the four-probe method performed in both liquid nitrogen and liquid helium without applied magnetic field. For both transport critical current and the joint resistivity measurements, the current source was a Sorenson P63, 0 – 1000 A DC power supply. The voltage was measured by a Keithley 2182A digital nanovoltmeter.

TEM samples were prepared by focused ion beam in the Thermal Fisher Scientific Helios G4 DualBeam scanning electron microscope . TEM was performed on the probe-aberration-corrected, cold emission JEOL JEM-ARM200cF at 200 kV with point resolution of 0.08 nm. The microstructures of the samples were imaged by high-angle-annular-dark field scanning transmission electron microscopy (HAADF-STEM) and annular-bright-field STEM (ABF-STEM).The general elemental chemical analysis in TEM was performed by energy dispersive spectroscopy (EDS) mapping. The oxygen content analysis in the REBCO layer was performed by electron energy loss spectroscopy spectrum imaging (EELS-SI). For EELS-SI, the electron probe size was 0.12 nm, and the energy resolution was 0.5 eV.

### III. Results

#### a. $I_c$ degradation due to heat treatment

We heat-treated SCS4050-AP samples for 2 hours at different temperatures in air. The $I_c$ of these samples are normalized to that of as-received samples and plotted against heat treatment temperature in Fig. 2(a). The data include both transport measurement at 77 K in self-field and magnetization measurement at 4.2 K in 8.5 T field perpendicular to *ab* plane. Evidently, considerable $I_c$ degradation occurs between 200 and 300 °C, which aggravates with increasing temperature. The relatively small difference in degradation between 77 K self-field transport data and 4.2 K 8.5 T magnetization data could be explained by the difference in flux pinning mechanisms at different temperature and field orientations [12]. $I_c$ of samples heat-treated at 300 °C for different durations are plotted in Fig. 2(b). As expected $I_c$ decreases monotonically with increasing heat-treatment time, as observed in Ref. [7].

It can be speculated that this thermal degradation is associated with oxygen out-diffusion from the REBCO layer. This oxygen out-diffusion causes oxygen deficiency in the REBCO layer, which results in partial or complete loss of its superconductivity. In order to prove this hypothesis, we use a simple one-dimensional oxygen out-diffusion model to fit our data. Here the effective diffusion coefficient $D$ can be written as,

$$D = D_0 \exp(-E_a/kT) \tag{1}$$

Where $D_0$ is a constant in m$^2$/s, $E_a$ is the activation energy, $k$ is the Boltzmann constant, and $T$ is the heat treatment temperature in kelvin. Following the Fick's law, the diffusion length $x$ which in our case corresponds to the thickness of the oxygen deficient layer can be written as

$$x = 2\sqrt{Dt} \qquad (2)$$

Where *t* is the heat treatment time in second.

Simply assuming that REBCO within the diffusion length *x* is no longer superconducting, $I_c$ can be written as,

$$I_c = I_{c0}(1 - x/d) \qquad (3)$$

Where $I_{c0}$ is the $I_c$ before heat treatment, *d* is the total REBCO layer thickness.

We use equations (1)-(3) to fit the experimental data and plotted as lines in Fig. 2(a) and 2(b). The fittings are quite satisfactory with $D_0$ = 2.5 x $10^{-6}$ m²/s, $E_a$ = 1.17 eV. So we have

$$D = 2.5 \times 10^{-6} \exp(-1.17 \text{ eV}/kT) \text{ m}^2/\text{s} \qquad (4)$$

In order to understand the role of Cu stabilizer in oxygen out-diffusion process, we heat treated an SF4050-AP sample (silver finish only) in air at 300 °C for 2 hours for comparison. In contrast to the severe $I_c$ degradation of SCS4050-AP sample with the same heat treatment, the SF4050-AP sample does not show significant $I_c$ degradation. When the SF4050-AP samples were heat treated in vacuum, however, its $I_c$ degraded similarly as observed in SCS4050-AP samples, as shown in Fig. 3. The fact that REBCO suffers similar $I_c$ degradation in vacuum with or without Cu stabilizer suggests that the oxygen out-diffusion process in SCS4050-AP sample is not hindered by the Cu stabilizer layer. More details of the oxygen out-diffusion process will be discussed in section IV.

### b. Effect of heat treatment on lap joint resistivity

The resistivity of joints made by heat treated REBCO were measured at both 4.2 and 77 K. Linear V-I traces were measured up to critical current for 77 K tests, and up to 200 A for 4.2 K tests. The joint resistivity is plotted as a function of temperature of 2.5 hours heat treatment (Fig. 4(a)) and heat treatment time at 200 °C (Fig. 4(b)). The solid lines are guides to the eye. Obviously joint resistivity increases monotonically with heat treatment temperature and time, which is in support of the oxygen out-diffusion model. It should be noted that even with a moderate heat treatment of 30 minutes at 200 °C, which is possible to be experienced by REBCO conductor in some coil fabrication processes, the 77 K joint resistivity doubles its original value. This increase in joint resistivity is likely linked to the existence of a thin layer of non-superconducting REBCO layer at REBCO/Ag interface as a result of the oxygen out-diffusion. At 4.2 K, the joint resistivity increment by heating is significantly less (Fig. 4(a)). This might be understood by $T_c$ distribution in the oxygen deficient region of the REBCO layer, where part of the region is superconducting at 4.2 K but not at 77 K. Since the oxygen out-diffusion causes both $I_c$ degradation and joint resistivity rise, we can correlate the joint resistivity with $I_c$ degradation ($I_{c0} - I_c$) as shown in Fig. 5. The strong monotonic correlation between joint resistivity and $I_c$ degradation further supports the thesis that joint resistivity is dominated by the resistive interface between REBCO and Ag, where the thickness of an oxygen deficient non-superconducting REBCO layer increases with heat treatment. This correlation also suggests that if $I_c$ is to be intentionally reduced by heat treatment, some level of increase in joint resistivity will be expected.

   c. **TEM investigation of heat treated REBCO**

We have shown that the one-dimensional oxygen out-diffusion model can explain the $I_c$ degradation and the joint resistivity rise very well. It is nonetheless highly desirable to find direct evidence of oxygen out-diffusion via microstructure analysis by cross-sectional TEM. The microstructures of the as-received, as well as samples heat treated for 2 hours at 195 °C, 250 °C and 300 °C are presented in Fig. 6 to Fig. 9.

The cross-section of the as-received sample is shown in Fig.6. The REBCO/Ag, and Ag/Cu interfaces are clean without unusual features from interfacial reactions. The average gain size of the Ag layer is 0.5 μm which remains approximately the same in the heat-treated samples. the REBCO/Ag interface is atomically sharp (Fig. 6(b) and 6(c)). The elemental maps (Fig. 6(d) to(g)) show uniform contrast with no additional features in the layers or at the interfaces.

When the sample is heat treated at 195 °C for 2 hours, an additional layer forms at the REBCO/Ag interface (Fig. 7(a) and 7(b)). The high resolution TEM image (Fig. 7(c)) shows that this layer contains small polycrystalline grains and the EDS composition maps confirms that it contains Cu and O (Fig. 7(d) to (g)). The thickness of this layer varies from 5 to 13 nm. The measured $d$-spacings (inset of Fig. 7(c)) from the diffraction spots in the fast Fourier transform (FFT) of one of the grains in this layer are 3.71 ± 0.2 Å and 3.00 ± 0.2 Å, which respectively correspond to {122} and {302} of the metastable Cu suboxide phase of $Cu_{64}O$. The formation of $Cu_{64}O$, a suboxide of Cu with very low O content, in this sample is conceivable. Because at 195 °C oxygen out-diffusion from REBCO is still weak, which favors formation of Cu-rich suboxide phase. This is comparable to the initial stage of Cu surface oxidation where the $Cu_{64}O$ phase was first identified [13]. The $d$-spacings measured from other gains (not shown) of this layer are 3.10 ± 0.2 Å and 2.47 ± 0.2 Å, corresponding to {110} and {111} of $Cu_2O$ phase respectively. Thus both $Cu_{64}O$ and $Cu_2O$ are found at the REBCO/Ag interface of this sample.

In the sample heat treated at 250 °C for 2 hours, Cu oxides are formed at both REBCO/Ag and Ag/Cu interfaces, as shown in Fig.8. It appears that Cu diffused through Ag grain boundaries to reach the REBCO/Ag interface. Meanwhile O out-diffused from the REBCO layer, reacted with the in-diffusing Cu on its path. So Cu oxidation occurred at both REBCO/Ag and Ag/Cu interfaces as well as at grain boundaries of the Ag layer which acts as fast Cu diffusion channels (Fig. 8(d)). The Cu oxide layer at the REBCO/Ag interface has a thickness of 5-20 nm.

At higher heat treatment temperature of 300 °C, more Cu diffused to the REBCO/Ag interface. Therefore more Cu oxides are formed at the REBCO/Ag interface and at Ag grain boundaries as shown in Fig. 9. The $d$-spacings measured from the selected area diffraction pattern of a Cu oxide region (Inset of Fig. 9(b)) are 3.00 ± 0.2 Å, 2.43 ± 0.2 Å and 2.09 ± 0.2 Å, which corresponds to {110}, {111} and {200} of $Cu_2O$. The thickness of this Cu oxide layer is 30 - 250 nm, much greater than that of 195 °C and 250 °C samples. It is also noticed that the thicker interfacial Cu oxide regions are near the Ag grain boundaries, demonstrating the key role Ag grain boundaries played in Cu in-diffusion.

EDS is not sufficiently sensitive to small changes in oxygen content, therefore not suitable for detecting oxygen deficiency in REBCO. EELS, on the other hand, has been used successfully to determine oxygen content in $YBa_2Cu_3O_{7-\delta}$ crystals [14]. The intensity of O-K pre-edge peak is correlated to the oxygen content $\delta$ in $YBa_2Cu_3O_{7-\delta}$ [15]. In our EELS-SI experiment, the focused electron beam was scanned over an area of 0.8 x 1.5 $\mu m^2$ with a step size of 32 nm in both directions. O-K EELS spectra were collected at each step with a dwell time of 0.2 seconds. The total of 26 spectra taken at the same distance from the REBCO/Ag interface are summed to enhance signal to noise ratio. For the as-received sample, in additional to the main O-K EELS edge, there is an pre-edge peak at 529 eV (Fig. 10(a)) indicating its near stoichiometry ($\delta \sim 0$). This is true for the entire REBCO layer. In contrast, for the 300 °C - 2 hours sample, the pre-edge peak disappears for the entire REBCO layer suggesting considerable loss of oxygen of the entire REBCO layer. Fig. 10(b) shows a few EELS spectra taken at different distances from the REBCO/Ag interface of the 250 °C - 2 hours sample. Evidently, the pre-edge peak which is associated with high oxygen content (low $\delta$ value) in $REBa_2Cu_3O_{7-\delta}$ gradually decreases in prominence with decreasing distance from the REBCO/Ag interface. This corresponds to an oxygen deficient REBCO layer near Ag/REBCO interface due to oxygen out-diffusion. The onset of O-K pre-edge disappearance is about 160 nm from the REBCO/Ag interface for the 250 °C - 2 hours sample. While for the 195 °C – 2 hours sample, the onset is at about 40 nm from the REBCO/Ag interface.

Attempts were made to quantify the oxygen content using the intensity ratio of the pre-edge peak to the main O-K edge following the approach described in ref [14]. But large uncertainties in peak areas make quantitative analysis very difficult. In addition, since it is not possible to predict the critical current of REBCO by its oxygen content alone, the correlation between our EELS spectra and critical current degradation remains qualitative.

## IV.     Discussions

The purpose of SCS4050 samples to be heat treated in air is to simulate REBCO coil fabrication processes which involve heating in air. In the temperature range of our interest, copper surface oxidation in air is relatively slow. For example, a study showed that a 300 °C- 2 hours oxidation in air only forms a 1.2 µm thick oxides layer [16]. With a 20 µm thick copper stabilizer, the oxygen cannot penetrate through the Cu stabilizer. So the oxygen present in the environment during heat treatment should not have any effect on the REBCO layer.

It is interesting to discover that for the non-stabilized sample SF4050 heat treated in air, $I_c$ degradation is insignificant, whereas these SF4050 samples heat-treated in vacuum has significant $I_c$ degradation that is comparable to that of the stabilized samples SCS4050. This can be explained as the follows. SF4050 only has a silver layer above REBCO. Silver is known to be permeable to oxygen [17]. When heat treated in air, oxygen out-diffusion does not occur. Because the oxygen partial pressure in air (about 0.2 bar) is slightly higher than equilibrium oxygen partial pressure of REBCO at that heat treatment temperature. According to references [18] and [19], the equilibrium oxygen partial pressure corresponding to the onset of REBCO decomposition is below 0.2 bar at 300 °C and decreases with temperature. Hence oxygen out-diffusion does not occur in SF4050 samples in air below 300 °C. This conclusion also suggests that silver stabilized conductor can be used to mitigate the $I_c$ thermal degradation for some special applications.

Our SCS4050 and SF4050 samples have very different original $I_c$ due to the difference in REBCO layer thickness ($J_c$ of these two samples were measured to be similar). As shown in Fig. 3, SCS4050 heat treated in air and SF4050 heat treated in vacuum suffer very similar $I_c$ degradation $\Delta I_c$ after the same heat treatment temperature and time. This suggests that in both cases the oxygen deficient layer have similar thickness which can be explained very well by our diffusion model. In terms of relative $I_c$ degradation, however, the thicker the REBCO layer, the lower the relative degradation for a given heat treatment.

Since the Ag layer does not act as an oxygen diffusion barrier, the $I_c$ degradation of SF4050 in vacuum is only limited by oxygen out-diffusion through REBCO crystal lattice. Since the copper stabilized SCS4050 samples degrade similarly, it is inferred that oxygen out-diffusion in SCS4050 is also controlled by oxygen diffusion through REBCO crystal lattice. The copper layer seems to act as an oxygen sink rather than a diffusion barrier by reacting with oxygen to form Cu-oxides as observed by TEM. The oxygen diffusion coefficient in REBCO single crystal is strongly anisotropic [20]-[24], i.e. its value along $ab$ plane $D_{ab}$ is up to 6 orders of magnitude greater than that along the c axis $D_c$. It was reported that at 400 C, $D_{ab}$ is ~$10^{-15}$ m$^2$/s while $D_c$ is ~$10^{-21}$ m$^2$/s [20]-[22]. In our samples, the oxygen diffusion direction is nominally along c axis. Our $D$ value of 5.2 x $10^{-15}$ m$^2$/s calculated by equation (4) at 400 °C, however, is comparable to $D_{ab}$ in the literature. Similar discrepancy was observed in [22] where SIMS studies showed that oxygen diffusion in the direction of c-axis of a YBCO film grown on Hastelloy C-276 is close to $D_{ab}$. The apparent high diffusion coefficient along c-axis might be due to the imperfections in the REBCO layer such as artificial pinning centers, threading dislocations etc. Likewise, our value of activation energy of 1.17 eV is within the range of 1.0 – 1.3 eV reported in [20]-[22] for $D_{ab}$, which is considerably lower than that for $D_c$ of 2.0 – 2.5 eV [21], [23]. It should also be noted that our model does not fit the data perfectly in Fig. 2. This seems to reflect the fact that the model we used is rather simplistic.

It is also interesting to observe the interaction between Cu, Ag, and O that diffused into Cu and Ag layers from the REBCO layer. It was initially speculated that O permeates through the Ag layer and react with Cu to form Cu oxides at the Cu/Ag interface. However, more Cu oxides are found at the REBCO/Ag interface as well as at the grain boundaries in the Ag layer. It seems that Cu diffuse into Ag layer and toward the Ag/REBCO interface and forms oxides whenever they meet out-diffusing oxygen. This behavior of Cu is puzzling, because the Cu-Ag interdiffusion is usually not very active at these temperatures. A cross-sectional TEM sample was made at the back side of the 300 °C - 2 hours heat-treated sample where there is a Cu/Ag interface but without REBCO. It shows no interdiffusion between Ag and Cu. This seems to imply that out-diffused oxygen plays an active role in facilitating Cu diffusion into and through the Ag layer.

One outcome of this work is that equation (1)-(4) can be used to predict $I_c$ degradation for a few cases of practical importance. Fig. 11 shows a heat treatment time-temperature diagram with two lines corresponding to oxygen out-diffusion length x = 0.1 μm and 1.0 μm, respectively. For a 1 μm thick REBCO layer, common in commercial coated conductor, the two lines divide the diagram into three regions: negligible degradation, significant degradation, and severe degradation. In the central region (shaded), some careful heat treatments could be performed to purposely lower critical current to a target value with a trade-off of increased joint resistivity. Another practical case of interest is the soldering process using either InSn (melting point 130 °C) or eutectic Pb37Sn63 (melting point 186 °C). Fig. 11 suggests that for a Pb37Sn63 soldering process at 200 °C, the soldering time can be up to 30 minutes without causing significant $I_c$ degradation or significant increase in joint resistivity. If the soldering process is at 220 °C, the soldering time needs to be under 10 minutes in order to avoid significant increase in joint resistivity. Another important case is the short high temperatures pulses experienced at localized hot spots during a magnet quench. In addition to the possible mechanical damage by thermal expansion mismatch, the $I_c$ degradation due to oxygen out-diffusion at these hot spots also needs to be considered. For example, if the hot spot is at 600 °C or above, it only takes a fraction of a second to severely degrade the conductor by oxygen out-diffusion.

V. **Conclusions**

We systematically studied the effect of heating on critical current and lab joint resistivity of REBCO coated conductor. It is found that $I_c$ decreases with heat treatment temperature and time. Meanwhile the resistivity of solder lap joints increases with heat treatment temperature and time. This effect is attributed to the oxygen out-diffusion from the REBCO layer causing degradation of its superconductivity. The effective oxygen out-diffusion coefficient is obtained by fitting the experiment data with a one-dimensional diffusion model. We found direct evidence of the oxygen out-diffusion in REBCO by EELS. The oxygen diffused out of the REBCO layer reacts with the copper in the stabilizer and forms Cu-oxides at both REBCO/Ag and Cu/Ag interfaces, as well as at grain boundaries of the Ag layer. This one-dimensional diffusion model is used to predict $I_c$ degradation of various scenarios of practical importance.

**Acknowledgement**

This work was supported in part by National Science Foundation Cooperative Agreement Nos. DMR-1644779 and DMR-1839796, and in part by the State of Florida.

**Captions**

Fig. 1 (a) Optical micrograph of a cross-section of SuperPower SCS4050-AP coated conductor, where the thin dark line above the Hastelloy C-276 is the REBCO layer. (b) HAADF-STEM image of the upper boxed region; (c) HAADF-STEM image of the Lower boxed region.

Fig. 2 (a) Normalized $I_c$ of SCS4050-AP samples heat-treated for 2 hours at different temperatures in air. Both 77 K self-field transport $I_c$ and 4.2 K magnetization $I_c$ measured in 8.5 T field perpendicular to ab plane are presented. (b) Normalized 77 K $I_c$ of SCS4050-AP samples heat-treated at 300 °C in air for different durations. The line in each plot is a simulation calculated by equations (1)-(4)

Fig. 3 $I_c$ measured at 77 K after heat treatment for 2 hours at different temperatures. A comparison of SF4050 (without Cu stabilizer layer) and SCS4050 (with Cu stabilizer layer).

Fig. 4 Lap joint resistivity of heat treated REBCO tapes. (a) joint resistivity versus heat treatment temperature for 2.5 hours of heat treatment, (b) joint resistivity versus time for 200 °C heat treatment.

Fig. 5 The correlation between joint resistivity of SCS4050-AP and its $I_c$ degradation $\Delta I_c$, both measured at 77 K. The solid line is a guide to the eye.

Fig. 6 Cross-sectional view of the as-received REBCO sample; (a) an SEM image showing the architecture of the tape; (b) atomic resolution HAADF-STEM image of the REBCO/Ag interface; (c) magnified REBCO/Ag interface showing the atomically sharp interface; (d) HAADF-STEM image of the EDS mapping region; (e) EDS Cu map. The apparent Cu background in the Ag layer is due to spurious signal from the copper sample holder; (f) EDS O map; (g) EDS Ag map.

Fig. 7 TEM of the sample heat treated at 195 °C for 2 hours. (a) BF-STEM image showing a thin white line at the REBCO/Ag interface; (b) magnified BF-STEM image of the REBCO/Ag interface. The extra phase is shown as white band indicated by the black arrows; (c) high resolution TEM of the REBCO/Ag interface

showing the crystalline grains of Cu oxide at the REBCO/Ag interface. Inset: FFT pattern from the Cu oxide grain; (d) HAADF-STEM image of the EDS mapping area. The dark band at the REBCO/Ag interface is the extra Cu oxide phase; (e) EDS Cu map; (f) EDS O map; (g) EDS Ag map.

Fig. 8 TEM of the sample heat treated at 250 °C for 2 hours; (a) HAADF-STEM image showing Cu oxides as dark contrast lines at both Ag/Cu and REBCO/Ag interfaces. (b) close-up of (a) at Ag/Cu interface, (c) close-up of (a) at REBCO/Ag interface. (d) EDS Cu+Ag map showing Cu oxides at Ag grain boundaries, (e) EDS Cu map of the REBCO/Ag interface. The apparent Cu background in the Ag layer is due to spurious signal from the copper sample holder (f) EDS O map of the same region. These EDS maps indicates that the diffused Cu is oxidized by oxygen.

Fig. 9 TEM of the sample heat treated at 300 °C for 2 hours. (a) HAADF-STEM image showing a dark line at the REBCO/Ag interface. There are also dark lines inside the Ag layer; (b) A close-up near the REBCO/Ag interface. The dark contrasts are indicated by a white arrow at the interface and by black arrows inside the Ag layer. Inset: selected area diffraction pattern from the interface; the bright spots are from the dark region and black spots are from Ag; (c) EDS Cu map. The apparent Cu background in the Ag layer is due to spurious signal from the copper sample holder; (d) EDS O map; (e) EDS Ag map; (f) EDS Ba map; These EDS maps verify that the extra phase at the REBCO interface is Cu oxides.

Fig. 10 EELS spectra of O-K edge taken from the REBCO layer. (a) typical O-K spectra from the as-received and 300 °C - 2 hours samples. The pre-edge peak is prominent in the spectrum of the as-received sample. It disappears in the spectrum of the 300 °C - 2 hours sample indicating its oxygen deficiency. (b) O-K spectra of the 250 °C – 2 hours sample taken at different distances from the REBCO/Ag interface (as labeled in nm).

Fig. 11 A diagram of heat treatment time-temperature indicating regions of different levels of degradation for a 1 μm REBCO layer. The lines which corresponds to two different oxygen out-diffusion lengths $x$ are calculated by equations (1) and (2). From left to right, the diagram is divided into three regions of negligible, significant and severe degradations. The melting point of Pb37Sn63 solder is shown in the diagram for reference.

J. Lu, et al, Fig. 1

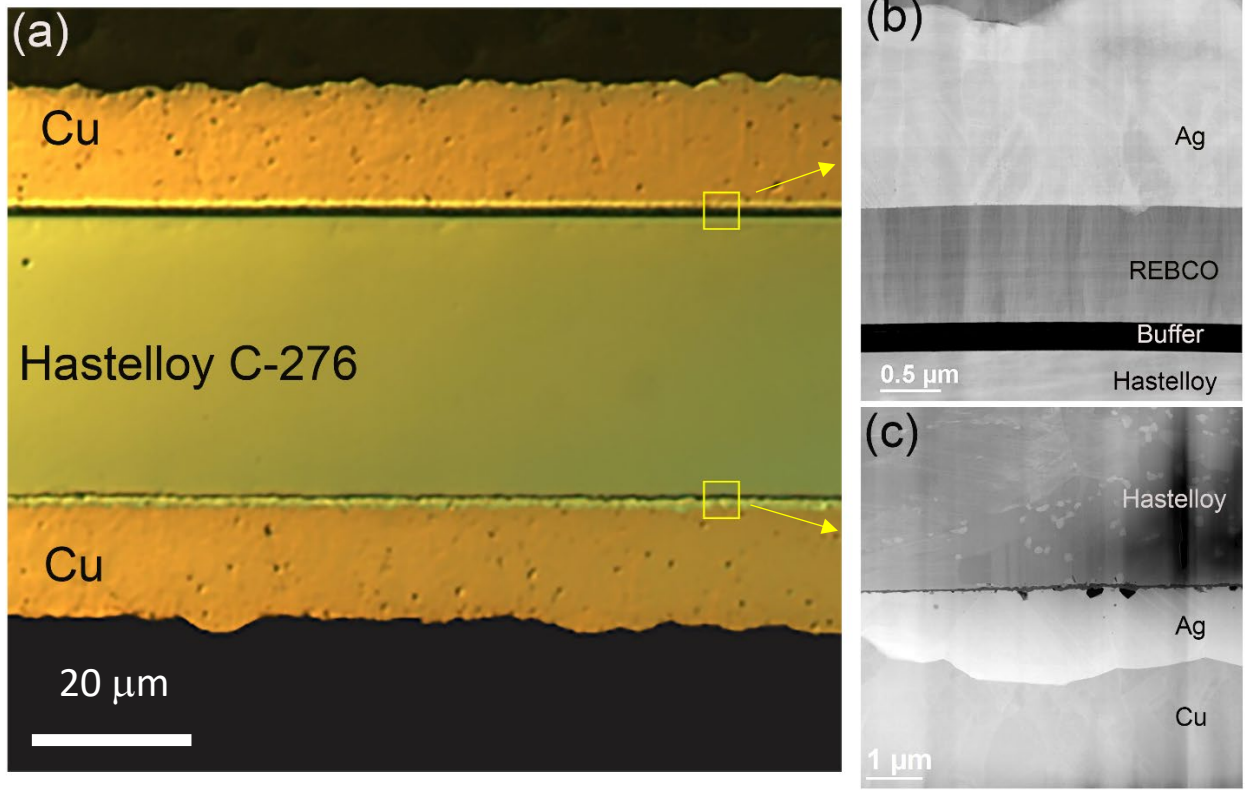

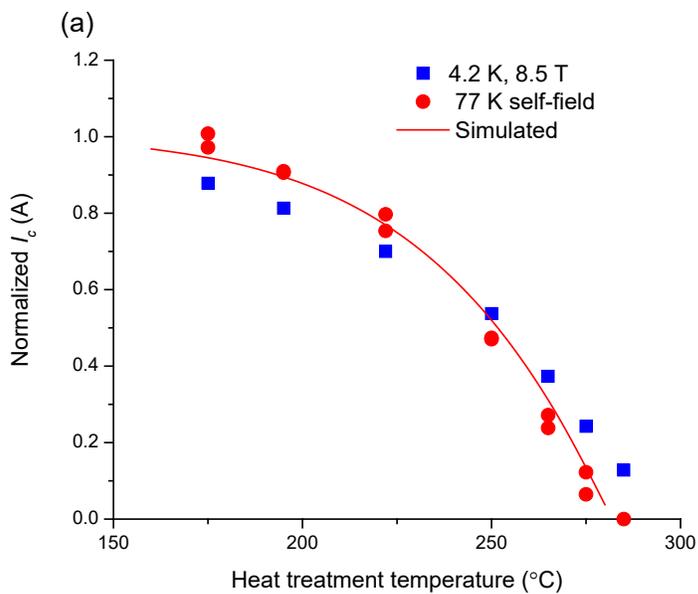

(a)

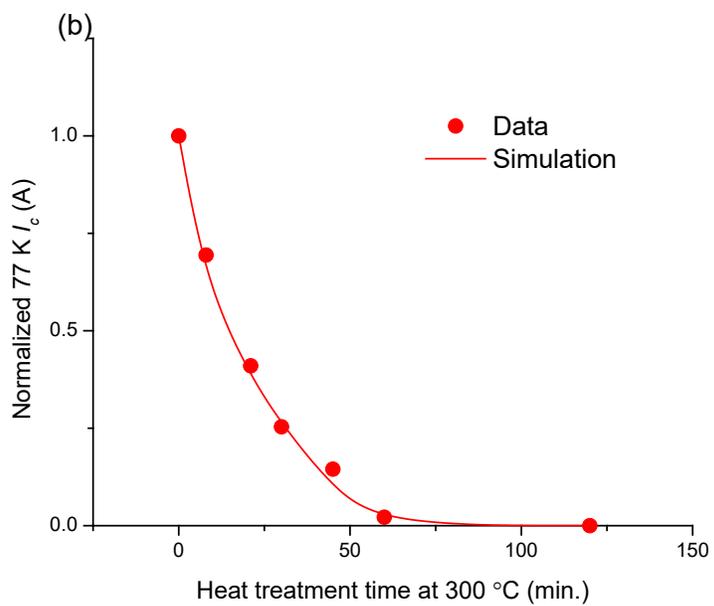

(b)

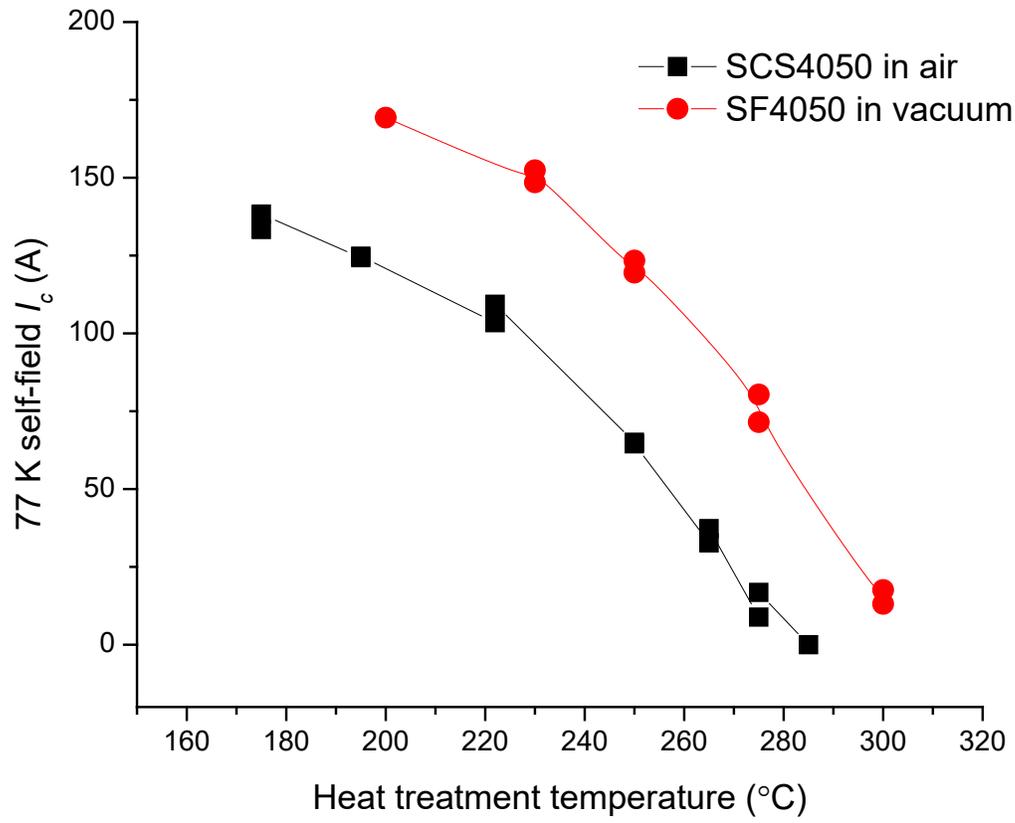

J. Lu, et al, Fig. 4

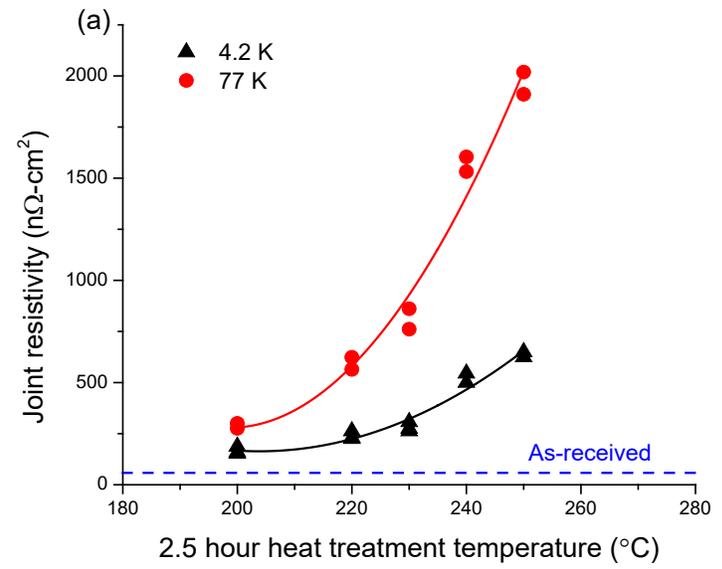

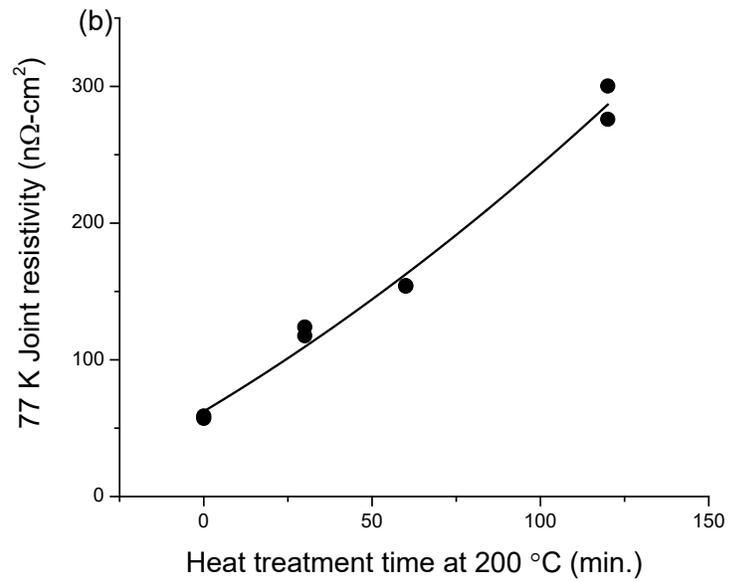

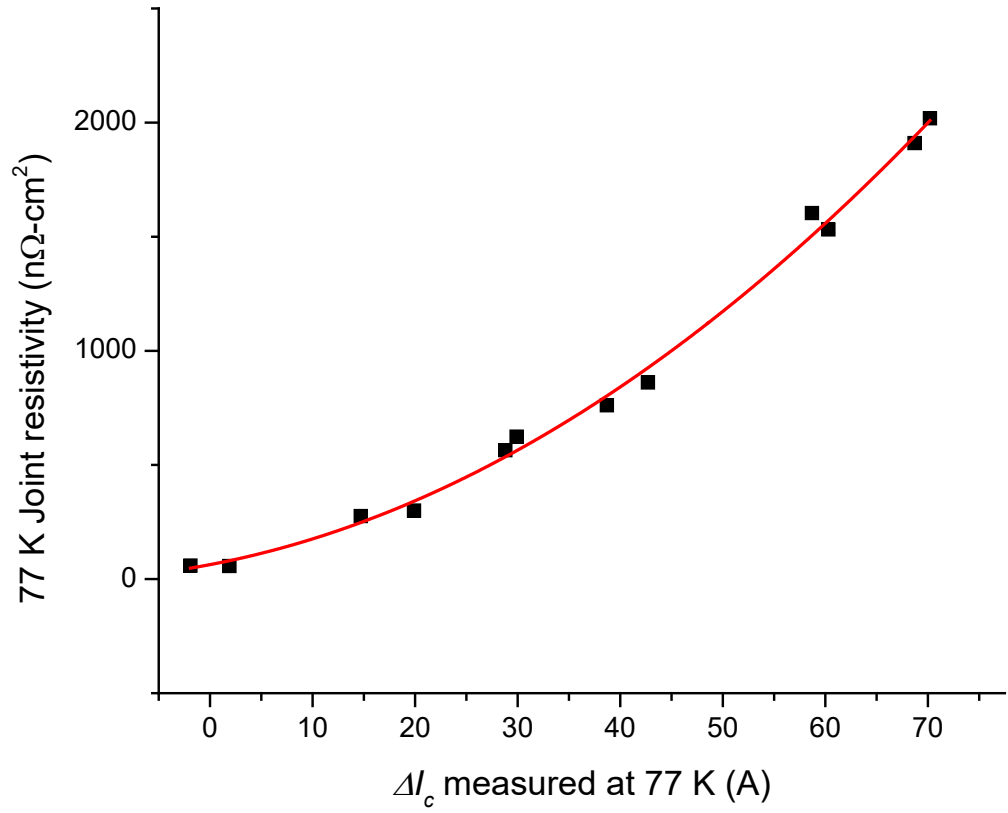

J. Lu, et al, Fig. 6

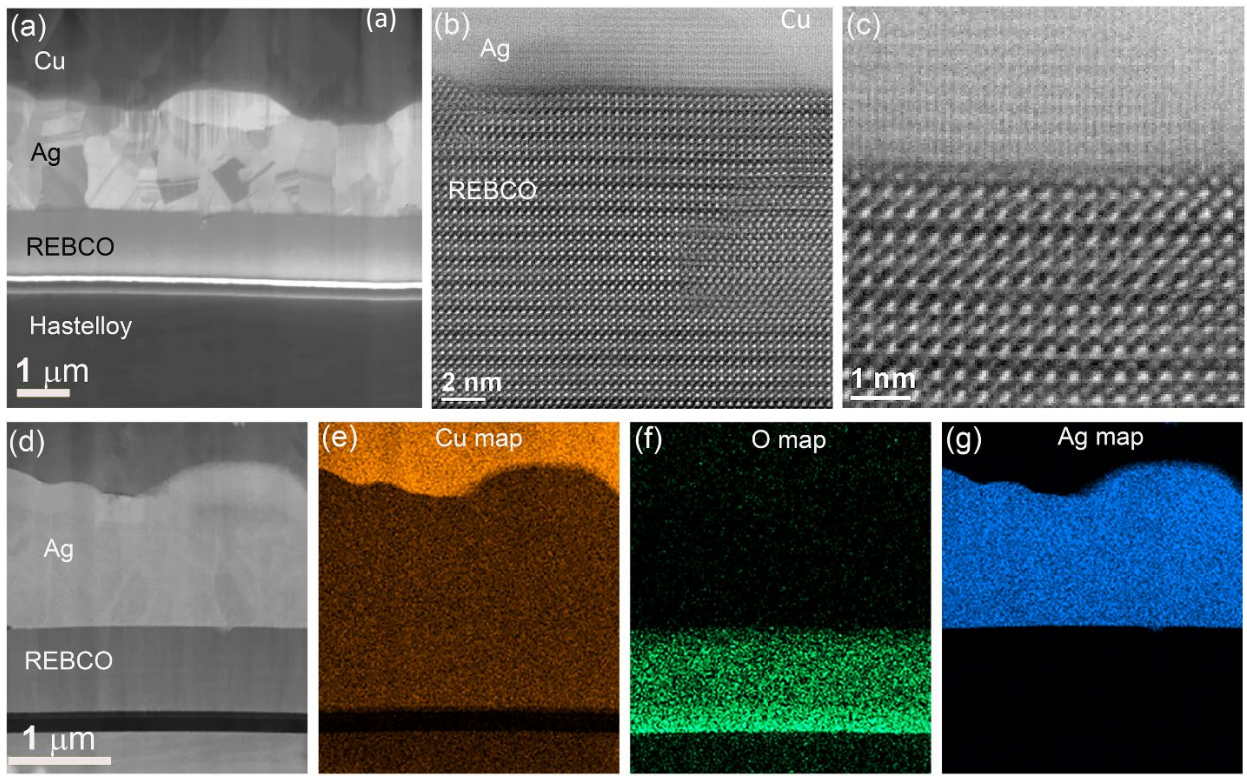

J. Lu, et al, Fig. 7

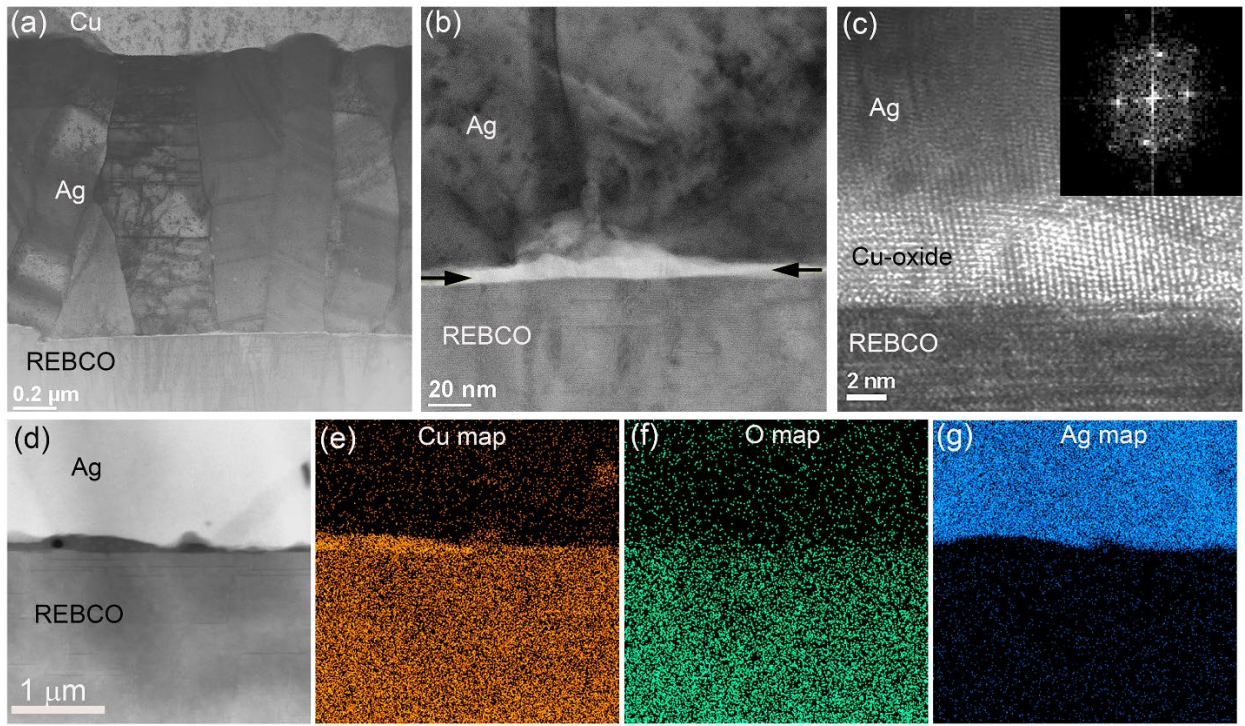

J. Lu, et al, Fig. 8

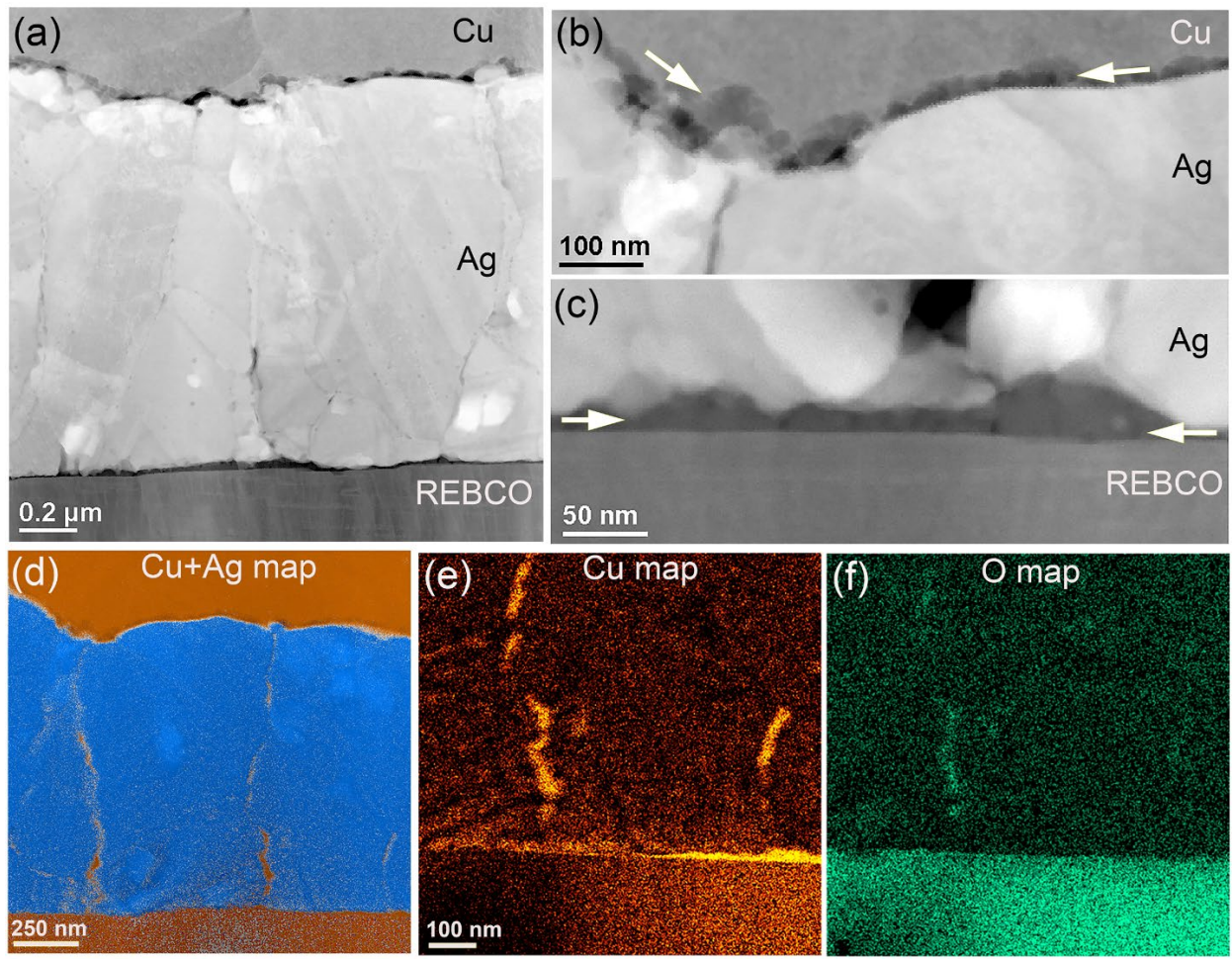

J. Lu, et al, Fig. 9

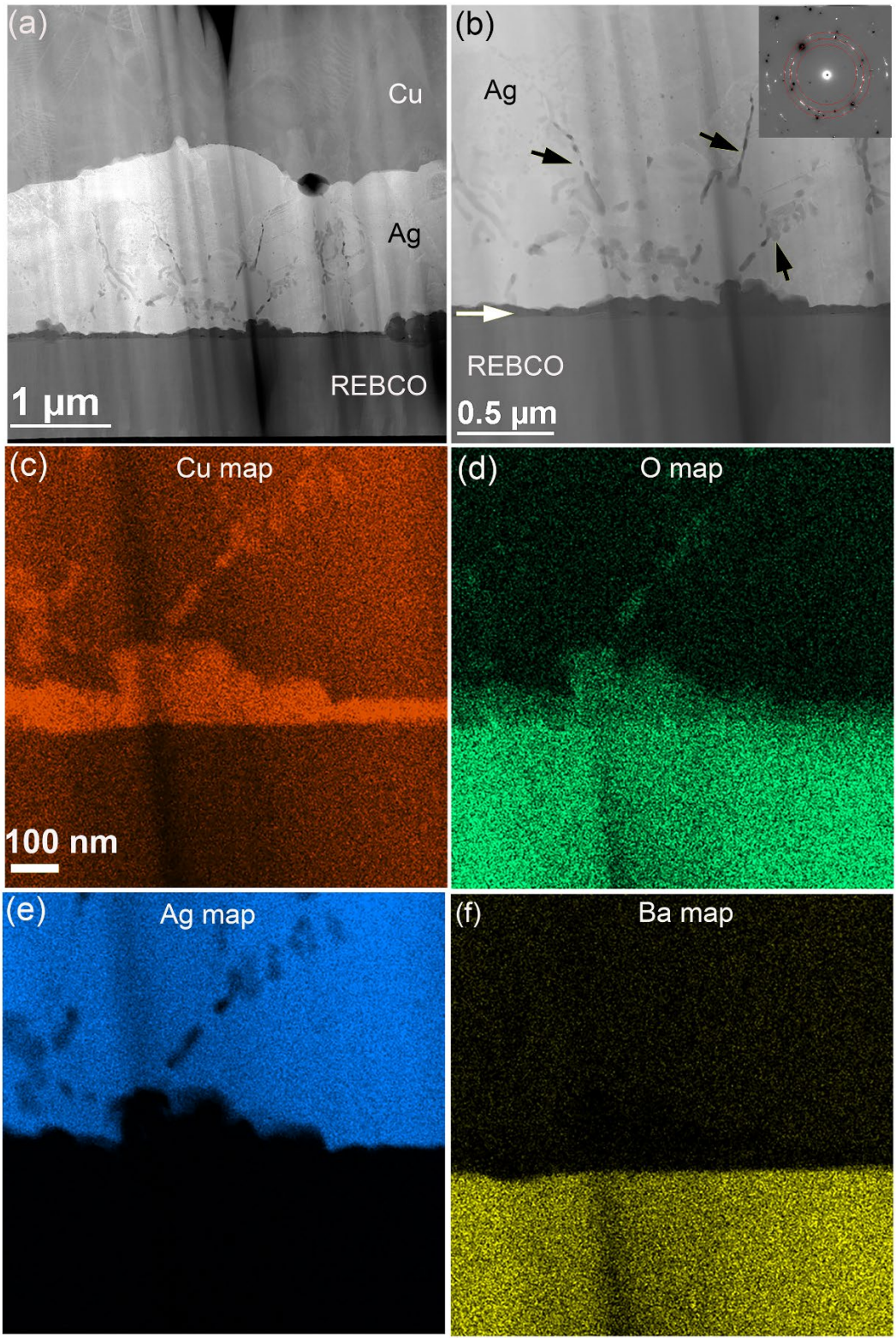

J. Lu, et al, Fig. 10

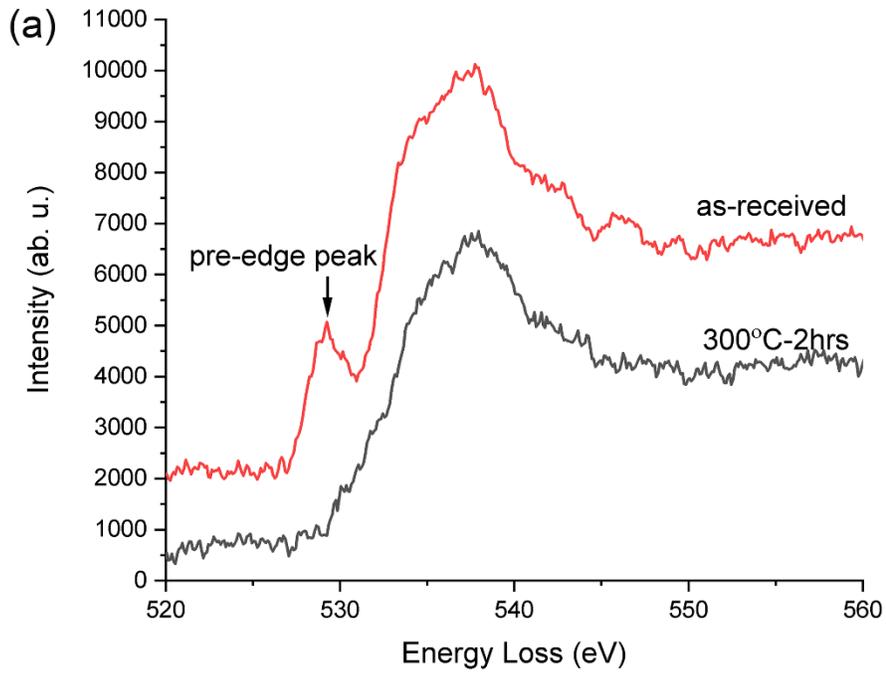

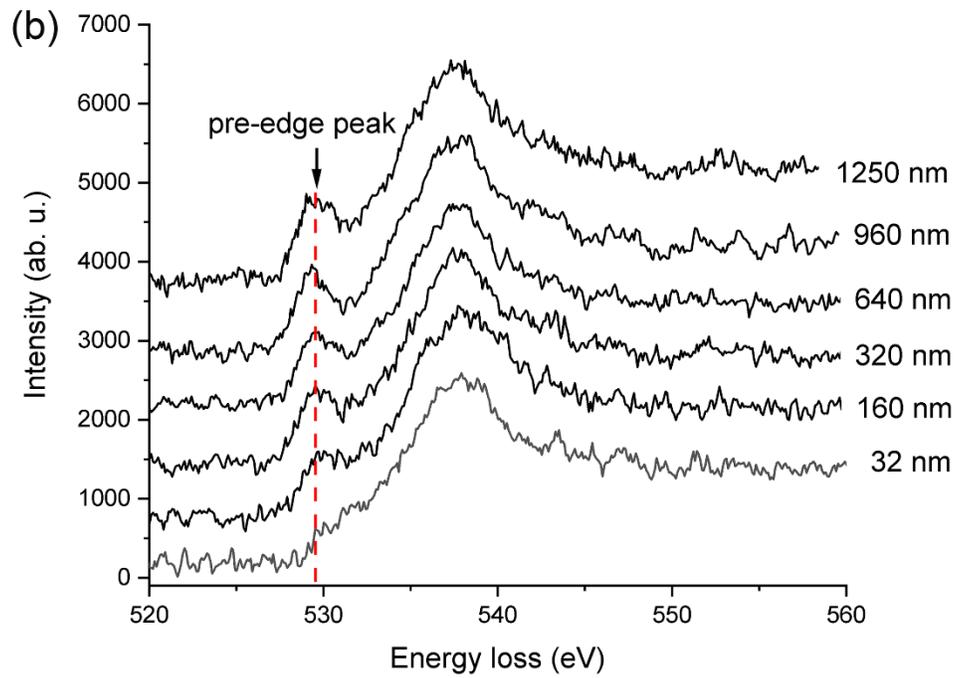

J. Lu, et al, Fig. 11

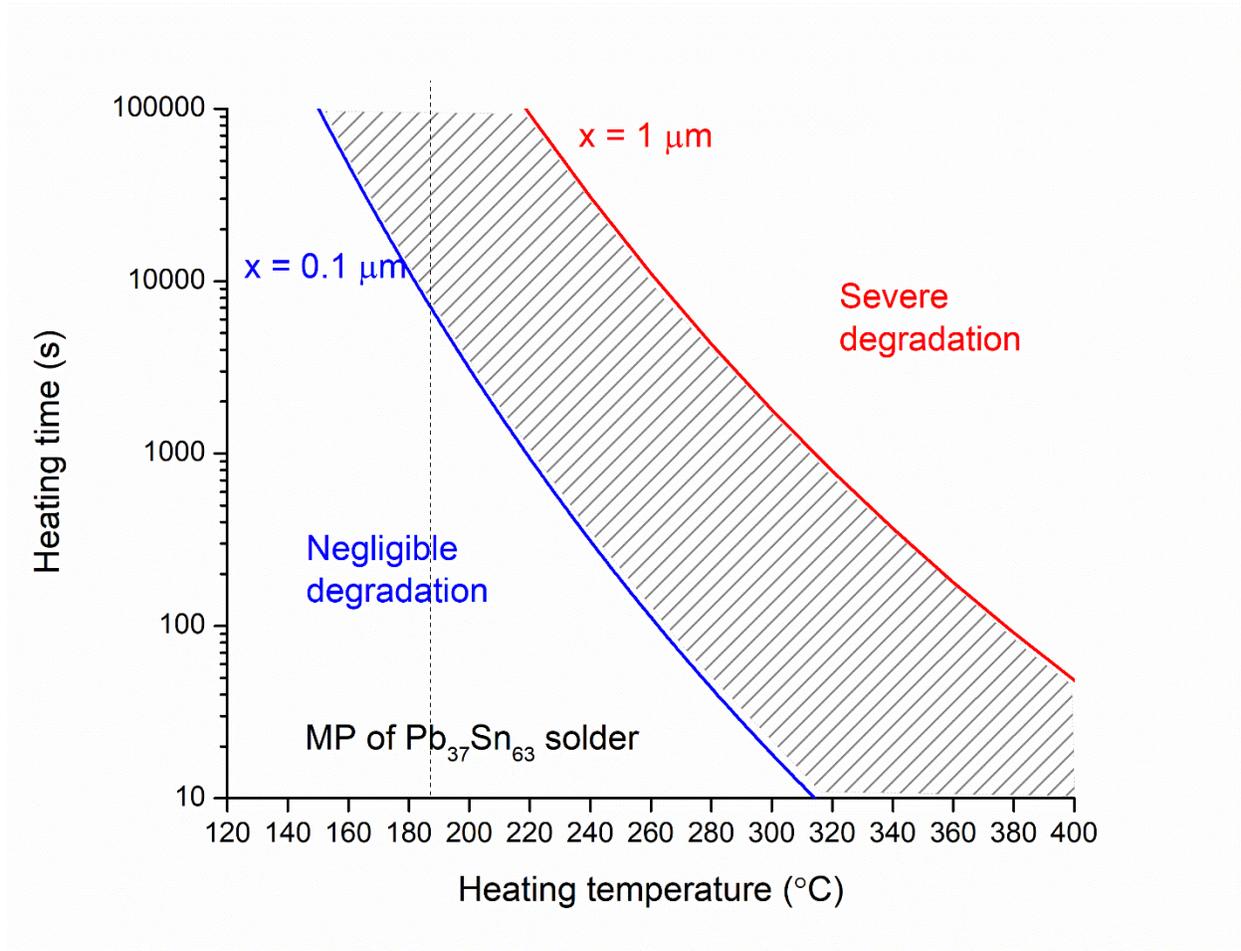